\documentstyle [12pt] {article}

\newcommand{\be}{\begin{equation}}
\newcommand{\ee}{\end{equation}}
\newcommand{\beqs}{\begin{eqnarray}}
\newcommand{\eeqs}{\end{eqnarray}}
\def\({\left (}
\def\){\right )}
\def\[{\left[}
\def\]{\right]}
\def\zlZ{{\zeta_2 \over \zeta_1}}
\def\zZl{{\zeta_1 \over \zeta_2}}
\def\zl{\zeta_1}
\def\zZ{\zeta_2}
\def\z{\zeta}
\def\d{\delta}
\def\D{\Delta}
\def\a{\alpha}
\def\b{\beta}

\def\na{\nabla}

\def\da{\dot{\alpha}}
\def\db{\dot{\beta}}
\def\pa{\partial}

\def\U{\Upsilon}
\def\tU{\tilde{\Upsilon}}
\def\bU{\bar{\Upsilon}}
\def\btU{\bar{\tilde{\Upsilon}}}

\def\th{\theta}
\def\ni{\noindent}
\def\nn{\nonumber}

\begin{document}

\begin{titlepage}

\begin{flushright}
\begin{tabular}{l}
ITP-SB-97-71    \\
hep-th/9712128  \\ 
December, 1997 
\end{tabular}
\end{flushright}

\vspace{8mm}
\begin{center}
{\Large \bf Feynman Rules in $N=2$ projective superspace III:}
 
\medskip
{\Large \bf Yang-Mills multiplet}

\vspace{4mm}
\vspace{16mm}

F.Gonzalez-Rey \footnote{email: glezrey@insti.physics.sunysb.edu},

\vspace{4mm}
Institute for Theoretical Physics  \\
State University of New York       \\
Stony Brook, N. Y. 11794-3840  \\

\vspace{20mm}

{\bf Abstract}
\end{center}

The kinetic action of the $N=2$ Yang-Mills vector multiplet can be
written in projective $N=2$ superspace using projective multiplets.
It is possible to perform a simple $N=2$ gauge fixing, which
translated to $N=1$ component language makes the kinetic terms of
gauge potentials invertible. After coupling the Yang-Mills multiplet
to unconstrained sources it is very simple to integrate out the gauge
fixed vector multiplet from the path integral of the free theory and
obtain the $N=2$ propagator. Its reduction to $N=1$ components agrees
with the propagators of the gauge fixed $N=1$ component
superfields. The coupling of Yang-Mills multiplets and hypermultiplets
in $N=2$ projective superspace allows us to define Feynman rules in
$N=2$ superspace for these two fields.

\vspace{35mm}

\end{titlepage}
\newpage
\setcounter{page}{1}
\pagestyle{plain}
\pagenumbering{arabic}
\renewcommand{\thefootnote}{\arabic{footnote}}
\setcounter{footnote}{0}

\section{Introduction}

 Quantization of $N=2$ superfields in $N=2$ superspace can be achieved
for multiplets living in certain subspaces of $N=2$ superspace. One 
such subspace is projective superspace \cite{martin_ulf}. Recently 
we have presented the Feynman rules for the quantization of massless 
hypermultiplets \cite{massless_hyper} and massive hypermultiplets
\cite{massive_hyper} living in this subspace (for an alternative 
description of $N=2$ supersymmetric systems see \cite{harmonic}).  
We consider now the $N=2$ Yang-Mills vector multiplet. 

 We briefly review the form of this multiplet in projective superspace
(the real tropical multiplet) as an infinite power series on the
projective complex coordinate. In the abelian case the $N=1$
components of the tropical multiplet can be simply related to the
prepotentials of the $N=1$ chiral spinor and chiral scalar contained
in the $N=2$ gauge field strength, plus pure gauge degrees of freedom
\cite{martin_ulf2}. In the nonabelian case the prepotentials we
mentioned are the ones corresponding to the kinetic action only,
and the field strengths have a highly nonlinear dependence on them.

 We propose a $N=2$ supersymmetric kinetic action for the projective 
Yang-Mills multiplet. In the abelian case it corresponds to the  
holomorphic $N=2$ gauge superpotential at tree level (in the literature
often called the $N=2$ prepotential). As usual, gauge fixing in 
the path integral is needed to be able to invert such kinetic terms. 
In $N=1$ superspace the prepotentials of the chiral spinor and chiral 
scalar appearing in the kinetic action each requires its own gauge 
fixing terms \cite{book}. We show that gauge fixing of the enlarged 
$N=2$ gauge symmetry \cite{martin_ulf2} reproduces the $N=1$ gauge 
fixing and we conjecture that it also introduces invertible kinetic 
terms for the pure gauge superfields. Following the same procedure as 
in \cite{massless_hyper}, we use the $N=1$ propagators of component 
superfields in the tropical multiplet to try and guess the form of a 
$N=2$ gauge propagator that contains them all.

 To justify our conjecture we quantize the action in $N=2$ superspace
by gauge fixing the $N=2$ real tropical multiplet as a whole. Once we
have an invertible kinetic term for this multiplet in projective
superspace, we can add the coupling to an unconstrained source and
integrate out the gauge potential from the free theory path integral
to find the $N=2$ propagator we guessed. In the nonabelian case we
have in addition interacting ghosts whose kinetic action is of the
same type as that of the hypermultiplet.

 Finally, we introduce suitable vertex factors describing the
interaction of the vector multiplet with the charged hypermultiplet. 
This multiplet is described by a complex superfield 
analytic in the projective complex coordinate \cite{massless_hyper}.
Self-interaction vertices for the nonabelian gauge multiplet are still
under investigation at the present time. We can give diagram
construction rules in $N=2$ superspace using the first type of vertices.

\section{Projective Superspace}

We briefly review the basic ideas of $N=2$ projective superspace. For 
a more complete review of $N=2$ projective superspace we 
refer the reader to \cite{martin_ulf},\cite{massless_hyper}. 

The algebra of $N=2$ supercovariant derivatives 
in four dimensions is\footnote{ We will use the notation and 
normalization conventions of 
\cite{book}; in particular we denote $D^2 = {1 \over 2} D^\a D_\a$
and $\Box = {1 \over 2} \pa^{\a\da} \pa_{\a\da}$.}

\be
 \{ D_{a \a} , D_{b \b} \} = 0 \; , \;\;\;
  \{ D_{a \a} , \bar{D}^b_{\db} \} = i \d^b_a \pa_{\a\db } \  .
\label{n2_algebra}
\ee

\ni
The projective subspace of $N=2$ superspace is parameterized 
by a complex coordinate $\zeta$, and it is spanned by 
the following projective supercovariant derivatives 
\cite{martin_ulf}

\beqs
 \na_\a (\z) & = & D_{1 \a} + \z D_{2 \a}   \\
 \bar{\na}_{\da} (\z) & = & \bar{D}^2_{\da} - \z \bar{D}^1_{\da} \ .
\label{proj_der}
\eeqs

 The conjugate of any object is constructed in this subspace by applying
the antipodal map to the complex coordinate stereo-graphically
projected onto the Riemann sphere, and composing it with complex 
conjugation back on the complex plane. To obtain the 
barred supercovariant derivate we conjugate the unbarred derivative and
we multiply by an additional factor $-\z$   

\be
- \z \overline{\na_\a (\z)} = \bar{\na}_{\da} 
 (\z) \ .
\ee

\ni
The projective supercovariant derivatives and the orthogonal combinations 

\be
 \D_\a (\z) = - D_{2 \a} + { 1 \over \z } D_{1 \a} \; , \;\;\;\;
 \bar{\D}_{\dot{\alpha}} (\z) = \bar{D}^1_{\dot{\alpha}} + { 1 \over \z }
 \bar{D}^2_{\dot{\alpha}} \ ,
\ee

\ni
constitute an alternative basis of spinor derivatives. They give the 
following algebra and identities

\beqs
& \{ \nabla (\z), \nabla (\z) \} = \{ \nabla (\z), \bar{\nabla} (\z)\}
 = \{ \Delta (\z), \Delta (\z)\} = \{ \Delta (\z) , \bar{\Delta} (\z) \} = 
 \{ \na (\z), \D (\z) \} = 0 &  \nn \\
& \{ \nabla_{\alpha} (\z), \bar{\Delta}_{\dot{\alpha}} (\z) \} = - 
 \{ \bar{\nabla}_{\dot{\alpha}} (\z) ,  \Delta_{\alpha} (\z) \} = 2i 
 \partial_{\alpha \dot{\alpha}} & \nn \\ 
& \{ \na_{\a} (\zl), \bar{\na}_{\da} (\zZ) \} = i (\zl - \zZ) 
 \pa_{\a \dot{\a} }&  \label{algebra_id} \\
& \na^2 (\zl) \na^2 (\zZ) = (\zl-\zZ)^2 (D_1)^2 (D_2)^2 & \nn \\
& \na^2 (\z) \D^2 (\z) = 4 (D_1)^2 (D_2)^2 \ . & \nn 
\eeqs

 For notational simplicity we will denote from now on 
$D_{1 \alpha} = D_\alpha , D_{2 \alpha} = Q_\alpha$. Superfields
living in $N=2$ projective superspace are annihilated by
the projective supercovariant derivatives (\ref{proj_der}). This
constraints can be rewritten as follows

\be
 D_\alpha \Upsilon = - \zeta \; Q_\alpha \Upsilon  \; \; , \; \; 
 \bar{Q}_{\dot{\alpha}} \Upsilon = \zeta \; \bar{D}_{\dot{\alpha}} 
 \Upsilon \ .
\label{proj_const}
\ee

\ni
Manifestly $N=2$ supersymmetric actions have the form 

\be 
 {1 \over 2 \pi i} \oint_C {d \zeta \over \zeta} \; d x \; D^2 \bar{D}^2 
 f(\Upsilon, \bar{\Upsilon}, \zeta) \ ,
\label{eq-action}
\ee

\noindent
where $C$ is a contour around some point of the complex plane 
that generically depends on the function $f(\Upsilon, \bar{\Upsilon},
\zeta)$.

 The superfields obeying (\ref{proj_const}) may be classified 
\cite{martin_ulf} as\footnote{ Throughout this paper we reserve the term 
multiplet to describe constrained superfields, while unconstrained
superfields with similar complex coordinate dependence are simply
called $O(k)$, rational, and analytic superfields.} : i) $O(k)$ 
multiplets, ii) rational multiplets iii) analytic multiplets.
The $O(k)$ multiplet can be expressed as a polynomial in 
$\zeta$ with powers ranging from $0$ to $k$. Rational multiplets are 
projective quotients of $O(k)$ superfields, and analytic 
multiplets are analytic in the coordinate $\z$ on some region of the
Riemann sphere. 

\noindent
For even $k$ we can impose a reality condition on the $O(k)$
multiplet. We refer to it as the real $O(2p)$ multiplet and we reserve 
the name $\eta$ for this field. The reality condition can be written 

\be
 \overline{\( \eta \over \z^p \)} = { \eta \over \z^p } \ ,
\label{eta_reality}
\ee

\noindent
or equivalently in terms of coefficient superfields

\be 
\eta_{2p-n} = (-)^{p-n} \bar{\eta}_n \ .
\ee

 The {\em arctic} multiplet is the limit $k \rightarrow \infty$ of 
the complex $O(k)$ multiplet. It is therefore
analytic in $\z$ around the north pole of the Riemman sphere

\be
\Upsilon = \sum_{n=0}^\infty \Upsilon_n \zeta^n \ .
\ee

\noindent 
Its conjugate ({\em antarctic}) superfield

\be
\bar{\Upsilon} = \sum_{n=0}^\infty \bar{\Upsilon}_n (- {1 \over \zeta})^n 
\ee

\noindent
is analytic around the south pole of the Riemann sphere. Similarly if 
we consider the self-conjugate superfield $\eta / \zeta^p$ the real 
{\em tropical} multiplet is the limit $p \rightarrow \infty$ 
of this multiplet

\be
V(\zeta, \bar{\zeta}) = \sum_{n= -\infty}^{+\infty} v_n \zeta^n \ .
\ee

\ni
It is analytic away from the polar regions and 
it contains a piece analytic around the north pole of the 
Riemann sphere (though not projective) and a piece analytic around the
south pole. The reality
condition in terms of its coefficient superfields is the following

\be 
v_{-n}= (-)^n \bar{v_n} \ .
\ee

 The constraints obeyed by multiplets living in projective superspace 
(\ref{proj_const}) can be written in terms of their coefficients

\be
 D_\alpha \Upsilon_{n+1} = - Q_\alpha \Upsilon_n  \;\; , \;\;
 \bar{D}_{\dot{\alpha}} \Upsilon_n = \bar{Q}_{\dot{\alpha}} 
 \Upsilon_{n+1} \ .
\ee

\noindent
Such constraints imply that the lowest order coefficient superfield 
of any multiplet is antichiral in $N=1$ superspace, and the next to 
lowest order is antilinear. The same 
constraints imply that the highest order coefficient superfield is
chiral in $N=1$ superspace and the next to highest order is linear.

\beqs
& D_\a \U_0 = 0 \;\; , \;\; D^2 \U_1 = 0 & \nn \\
& \bar{D}_{\dot{\alpha}} \bar{\U}_0 = 0 \;\; , \;\; 
 \bar{D}^2 \bar{\U}_1 = 0 & .
\label{n1_constr}
\eeqs

\ni
In the case of a complex $O(k)$ hypermultiplets its highest
and lowest order superfields are not conjugate to each other, and
the complex multiplet describes twice as many physical degrees of freedom as
the real one \cite{massless_hyper}.

 In the case of the real projective multiplet there is no lowest or
highest order coefficient, and therefore none of the
coefficient superfields is constrained in $N=1$ superspace.

\section{Kinetic Yang-Mills action in $N=2$ superspace}

 The minimal action of the $N=2$ Yang-Mills multiplet is well known

\be
S = {1 \over 2} \; Tr \; \( \int {d x D^2 Q^2} W W + 
 \int {d x \bar{D}^2 \bar{Q}^2} \bar{W} \bar{W} \) \ ,
\label{gauge_action}
\ee

\noindent
where the $N=2$ superfield strength $W$ is a covariantly chiral scalar
proportional to the anticommutator of gauge covariantized $N=2$ spinor 
derivatives 

\be
 \{ {\cal D}_\alpha , {\cal Q}_\beta \} = i C_{\alpha \beta} 
 \bar{W} \ .
\ee

\ni
Its expansion in the Grassmann coordinate of the second supersymmetry 
gives the N=1 covariantly chiral field strengths

\be
 W_{| \th_2 = 0} = \Phi \;\;\; ,\;\;\; {\cal Q}_\a W_{| \th_2 = 0} =
 - W_\a \ .
\ee

\ni
The action expressed in terms of this fields is

\be
S =  \int {d x D^2 \bar{D}^2} \; Tr \; \bar{\Phi} \Phi + {1 \over 2} 
 \( \int {d x D^2} \; Tr \; {W^\alpha W_\alpha \over 2} + 
 \int {d x \bar{D}^2} \; Tr \;
 { \bar{W}^{\dot{\alpha}} \bar{W}_{\dot{\alpha}} \over 2} \) \ . 
\label{n1_actionn}
\ee

\ni

The term quadratic in the $N=1$ covariantly chiral field $\Phi$ and
its conjugate contains the interactions of an ordinary chiral scalar
with the other degrees of freedom in the gauge multiplet. This is made
manifest by using the gauge chiral representation of the gauge
covariantized $N=1$ spinor derivatives \cite{book}. The barred derivatives 
annihilate ordinary chiral fields while the unbarred ones annihilate 
covariantly antichiral fields defined in terms of ordinary antichiral 
superfields and a real gauge prepotential $v$

\be
 \int {d x D^2 \bar{D}^2} \; Tr \; \bar{\Phi} \Phi = 
 \int {d x D^2 \bar{D}^2} \; Tr \; e^v \bar{\phi}  e^{-v} \phi \ . 
\ee

\ni
The gauge superfield $v$ is a prepotential for the $N=1$ chiral
spinor field strength   

\be
 W_\alpha = i \bar{D}^2 ( e^{-v} D_\alpha e^{v} ) \ ,
\ee

\ni
and similarly the chiral field $\phi$ can also be defined by a complex
prepotential 

\be
\phi = \bar{D}^2 \bar{\psi} \ .
\ee

\ni
The kinetic part of the action (\ref{n1_actionn}) written in terms of
$N=1$ prepotentials is then

\be
S_0 = \int d^4x D^2 \bar{D}^2 \;Tr \; \( \psi D^2 \bar{D}^2 \bar{\psi} + 
 {1 \over 2} v D^\alpha \bar{D}^2 D_\alpha v \) \ .
\label{n1_gauge_kin}
\ee

\ni
In the abelian theory this action is also the full gauge action because
there are no self-interactions

 This well known description of the gauge multiplet can be related to 
a real tropical multiplet $V(\zeta)$ that we will call the projective 
vector multiplet. 
For the abelian multiplet the relation among component fields of both
descriptions is very simple and direct. For the nonabelian multiplet 
the relation is very nonlinear, but we can still formulate the theory 
in terms of projective vector multiplets.

 To understand the projective superspace description of the gauge
multiplet and write the kinetic action (\ref{n1_gauge_kin}) using 
real tropical multiplets, we consider the $N=2$ supersymmetric 
interaction of a real tropical multiplet and a complex (ant)arctic 
hypermultiplet \cite{martin_ulf2} in the (anti)fundamental 
representation of the gauge group

\be 
S_\U = \int d x d^4 \th \oint {d\zeta \over 2 \pi i \zeta} \; \; 
 \bar{\Upsilon} e^{V} \Upsilon  \ .
\label{eq-Lagrangian2}
\ee

\ni 
This action is invariant under gauge transformations

\be
 \bU' = \overline{\( \U'\)} = \bU e^{-i \bar{\Lambda}} 
 \;\;\; , \;\;\; (e^V)' = e^{i \bar{\Lambda}} e^V e^{-i \Lambda}  
 \;\;\; , \;\;\;  \U' = e^{i \Lambda} \U \ ,
\ee

\ni
where the gauge parameter is an (ant)arctic multiplet. This guarantees
that the transformed hypermultiplet is also (ant)arctic. 
The infinitesimal {\em abelian} transformation of the real tropical 
multiplet in terms of $\z$-coefficient superfields \cite{martin_ulf2} 
is 

\be
\delta V = i (\bar{\Lambda} - \Lambda) \longrightarrow \delta v_0 =
 i(\bar{\lambda}_0 - \lambda_0) \;\;\; , \;\;\; \delta v_n = 
 - i \lambda_n \ .
\label{vector_transf}
\ee

\ni 
Since $\lambda_0$ is antichiral and $\lambda_1$ is antilinear, while
higher order coefficients are unconstrained, the gauge transformation
can be used to identify the physical degrees of freedom in the real
tropical multiplet \cite{martin_ulf2}. First we put the real tropical
multiplet in a gauge where it becomes a real $O(2)$ multiplet
by setting the components $v_n = 0 \; \forall \, n \neq -1,0,1$. We
can further gauge away all of $v_1$ except for the antichiral piece
$D^2 v_1$ and correspondingly keep the chiral piece in $v_{-1}$, taking
this $O(2)$ gauge multiplet to a Lindstr\"{o}m-Ro\v{c}ek gauge.
Finally, we can put the coefficient $v_0$ in a Wess-Zumino gauge, and
then we have isolated the physical degrees of freedom contained in $V
(\z)$. This suggests \cite{martin_ulf2} that $i v_{-1} = \bar{\psi}$ is a
prepotential for the chiral scalar gauge field strength, $i v_1 = \psi$ is
a prepotential for the antichiral scalar, and $v_0 = v$ is the usual
$N=1$ prepotential of the chiral spinor gauge field strength $W_\a = -
i Q_\a \bar{D}^2 v_{-1} = i \bar{D}^2 D_\a v_0$. All other coefficient
superfields in $V$ are gauge degrees of freedom.

If the real tropical multiplet is Lie algebra valued, the
corresponding {\em nonabelian} infinitesimal transformation
is highly nonlinear

\be
\d V = L_{V \over 2} \left[ -i \( \bar{\Lambda} + \Lambda \) + coth 
  L_{V \over 2} i \( \bar{\Lambda} - \Lambda \) \right] \ ,
\ee

\ni
where the Lie derivative is defined as the commutator \cite{book}

\be
 L_{Y} X= [Y, X] \ . 
\ee

\ni
The individual components in $V$ transform in a complicated way, but 
we can see that it is possible to put the nonabelian
real tropical multiplet in an $O(2)$ gauge by noticing that the most
general real tropical multiplet can be written as a gauge transformed
$O(2)$ multiplet

\be
 V = V^{O(2)} + L_{V^{O(2)} \over 2} \left[ -i \( \bar{\Lambda} + 
   \Lambda \) + coth L_{V^{O(2)} \over 2} i \( \bar{\Lambda} - \Lambda \) 
   \right] \ .
\ee

\ni
The linearized transformation is of the form (\ref{vector_transf}),
giving a most general real tropical multiplet. The nonlinear
corrections do not change this condition. 

The projective superspace description of the gauge multiplet can be
used to construct an explicit representation of gauge covariantized 
spinor derivatives \cite{martin_ulf2}. We split the exponential of 
the real tropical 
multiplet into a part analytic around the north pole of the Riemann 
sphere and a part analytic around the south pole\footnote{ In the 
abelian case the exponents correspond to the polar pieces of the 
tropical multiplet $V= V_+ + V_-$, but for the nonabelian multiplet this
is not true.}  

\be
 e^{V} = e^{V_+} e^{V_-} \; , \;\;
 e^{V_-} = \overline{ \( e^{V_+} \) } \ .
\ee

\ni
Using the fact that the vector multiplet is a projective multiplet 
$ \( \na_\a e^V \) = 0$, we can see that the projective spinor derivatives

\be
 \tilde{\na}_\a = e^{V_+} \na_\a e^{- V_+} = e^{- V_-} \na_\a e^{ V_-}
\ee 

\ni
and
 
\be
 \tilde{\bar{\na}}_{\da} = e^{ V_+} \bar{\na}_{\da} e^{- V_+}=
 e^{- V_-} \bar{\na}_{\da} e^{ V_-} 
\ee

\ni
annihilate a {\em covariantly} projective (ant)arctic multiplet

\be
 \tilde{\U}^i = \(e^{V_+}\)^i_{\ j} \U^j \; , \;\;\; 
 \tilde{\bar{\U}}_i = \bar{\U}_j \( e^{V_-} \)_{\ i}^j  \ .
\ee

\ni
In the abelian case it is very easy to evaluate the anticommutator 
of such gauge covariantized spinor derivatives  

\be
 \{ {\cal D}_\a , {\cal Q}_\b \} = C_{\a \b} D^2 v_1 \; , \; \;
 \{ \bar{\cal D}_{\da} , \bar{\cal Q}_{\db} \} = C_{\da \db} 
 \bar{D}^2 v_{-1} \ .
\label{central_charge}
\ee

\ni
They are proportional to the $N=2$ abelian gauge field strengths 
$\bar{W} = i D^2 v_1$ and its conjugate. In the nonabelian case 
the anticommutator also defines the gauge field strength, although its
explicit form is highly nonlinear in the $\z$-coefficient superfields 
and (\ref{central_charge}) gives only the lowest order terms. 

 To see that the action (\ref{eq-Lagrangian2}) describes 
hypermultiplets interacting with a gauge multiplet we rewrite this
action using covariantly projective hypermultiplets

\be
S_\U = \int d x d^4 \th \oint {d\zeta \over 2 \pi i \zeta} \; \; 
 \bU e^{V} \U = \int d x d^4 \th \oint {d\zeta \over 2 \pi i \zeta} 
 \; \; \btU \tU \ , 
\label{cov_proj}
\ee

\ni
and we perform 
the duality  that (in the hypermultiplet free theory)
exchanges the complex linear field of the (ant)arctic multiplet by a
chiral one \cite{massless_hyper}. This duality relates the off-shell $N=2$
description of the hypermultiplet to the traditional on-shell
realization. The algebra (\ref{central_charge}) induces a modified 
$N=1$ linearity constraint on $\btU_1$ \cite{massive_hyper}. We can
impose this constraint using a Lagrange multiplier in the conjugate 
fundamental representation of the gauge group

\be
S_\U = \int d x \; d^4 \th \; \( \btU_0 \tU_0 - \btU_1 \tU_1 + 
 \btU_2 \tU_2 + \dots + \tilde{Y} (\bar{\cal D}^2 \btU_1 - i W \btU_0) 
 + \bar{\tilde{Y}} ({\cal D}^2 S + i \bar{W} \tU_0) \) \ .
\ee

\ni
We can integrate out the unconstrained field $\tU_1$ in the path integral
of the theory. The dualization gives the action of two $N=1$
covariantly chiral scalars $\bar{\tilde{\U}}_0$ and 
$\bar{\cal D}^2 \tilde{Y}$ (in the fundamental and antifundamental
representation respectively) interacting with a chiral gauge scalar 
$W|= \Phi$. In addition we have auxiliary fields that decouple

\beqs
S_{dual} & = & \int d x \; D^2 \bar{D}^2 \; \; ( 
 \tilde{\U}_0 \bar{\tilde{\U}}_0 + 
 {\cal D}^2 \bar{\tilde{Y}} \bar{\cal D}^2 \tilde{Y} + \dots ) \nn \\ 
& & - i \int d x \; D^2 ( \bar{\cal D}^2 \tilde{Y} \Phi \bar{\tilde{\U}}_0) 
 + i \int d x \; \bar{D}^2 ({\cal D}^2 \bar{\tilde{Y}} \bar{\Phi} 
 \tilde{\U}_0 ) \ .    
\label{dual_action}
\eeqs

\ni
It is possible to rewrite this dual action in terms of ordinary chiral
fields by going to the gauge chiral representation of the gauge 
covariantized derivatives 

\beqs
S_{dual} & = &\int d x \; d^4 \th \; ( \U_0 e^v
 \bar{\U}_0 + D^2 \bar{Y} e^{-v} \bar{D}^2 Y   
 + \dots )  \nn  \\
& & - i \int d x \; d^2 \th ( \bar{D}^2 Y \phi \bar{\U}_0 )
 + i \int d x \; d^2 \bar{\th} \; ( D^2 \bar{Y} \bar{\phi} \U_0) \ . 
\eeqs

 This well known $N=1$ formulation of the charged hypermultiplet gives
1-loop logarithmic divergences \cite{n1} proportional to the gauge kinetic
action (\ref{n1_gauge_kin}). In our formulation we can compute
analogous $N=1$ diagrams \cite{eff_action} with the component 
field interactions in (\ref{eq-Lagrangian2}). Combining all the 
diagrams with external potentials $v_i$ and hypermultiplets running 
in the loop, we find that the final logarithmic divergence is indeed 
proportional to (\ref{n1_gauge_kin}).

 We also have the tools to compute these 1-loop divergences directly 
in $N=2$ superspace \cite{eff_action}: we use the hypermultiplet 
$N=2$ propagator given in ref. \cite{massless_hyper} and the 
projective superspace interactions in (\ref{eq-Lagrangian2}). Just as 
in the $N=1$ calculation, the logarithmic divergences must be 
proportional to the gauge kinetic action and induce a wave function
renormalization.

 The only nonvanishing logarithmic divergence we find comes from the
two point function $\langle V(1) V(2) \rangle$. It is an integral of a
function local in $N=2$ superspace but nonlocal in the complex
coordinate $\z$: it involves two complex contour integrals on
overlapping contours around the origin of the complex plane. Its
projection to the chiral and antichiral $N=2$ subspaces can be
trivially integrated on the complex coordinates

\beqs
 S_0 & = & - {Tr \over 2} \int d x d^8 \th \oint {d \zl \over 2 \pi i}
 {d\zZ \over 2 \pi i} { V(\zl) V(\zZ) \over (\zl - \zZ)^2 } 
 \label{n2_vector_action}  \\
& = & - {Tr \over 2} \int d x D^2 Q^2 \oint {d\zeta_1 \over 2 \pi i}  
 {d\zeta_2 \over 2 \pi i} { \bar{\D}^2_1 V(1) \bar{\na}^2_1 V(2) 
 \over 4 (\zeta_1- \zeta_2)^2 }   \nn  \\
& = & - {Tr \over 2} \int d x D^2 Q^2 \oint {d\zeta_1 \over 2 \pi i}  
 {d\zeta_2 \over 2 \pi i} \bar{D}^2 V(1) \bar{D}^2 V(2) \nn \\
& = & - {Tr \over 2} \int d x D^2 Q^2 \( \bar{D}^2 v_{-1} 
 \bar{D}^2 v_{-1} \) \nn \ .
\eeqs

\ni
In the abelian case this $N=2$ superpotential is the tree level gauge 
action (\ref{gauge_action}). We can write this expression in $N=1$ 
superspace 

\beqs
S_0 & = & - {1 \over 2} \int dx D^2 Q^2 \; Tr \; ( \bar{D}^2 v_{-1} 
 \bar{D}^2 v_{-1}) \\
& = & - {1 \over 2} \int d x D^2 \; Tr \; 
 ( 2 \bar{D}^2 Q^2 v_{-1} \bar{D}^2 v_{-1} 
 + \bar{D}^2 Q^\a v_{-1} \bar{D}^2 Q_\a v_{-1} ) \nn  \\
& = & - \int d x D^2 \bar{D}^2 \; Tr \; ( D^2 v_1 \bar{D}^2 v_{-1} - 
 {1 \over 2} v_0 D \bar{D}^2 D v_0 ) \ , \nn  
\eeqs  

\ni
and we find the kinetic action (\ref{n1_gauge_kin}) after identifying 
again $i v_1 = \psi, \ v_0 = v$.

\section{$N=1$ gauge fixing}

 It is well known that the $N=1$ action (\ref{n1_gauge_kin}) does not have
invertible kinetic terms and the system needs gauge fixing to remove 
the gauge group volume from the path integral. Suitable gauge fixing
conditions for the $N=1$ prepotentials are \cite{book} 

\be
D^2 v = 0 = \bar{D}^2 v \;\;\; , \;\;\; D_\alpha \bar{\psi} = 0 = 
 \bar{D}_{\dot{\alpha}} \psi \ .
\label{n=1_gauge_fix}
\ee

\ni
This gauge fixing is imposed by inserting unity in the path integral
as the product of a functional Dirac delta times the inverse
Faddeev-Popov determinant. The determinant can be written as a
functional integral of an exponential. The exponent is just the 
gauge fixing function evaluated on anticommuting unconstrained scalar 
and spinorial ghosts \cite{book}. 
These unconstrained prepotential ghosts can be traded for
anticommuting chiral and complex linear field strength ghosts,
and their kinetic term is  

\be
S_{FP} = \int d x d^4 \th \( b D^2 \bar{D}^2 \bar{c} + 
 b^\a D_\a \bar{D}^{\da} \bar{c}_{\da} + c.c. \) = 
 \int d x d^4 \th \(D^2 b \bar{D}^2 \bar{c} - D^\a  b_\a 
 \bar{D}^{\da} \bar{c}_{\da} + c.c. \) \ .
\label{n1_ghosts}
\ee

 With convenient gauge fix averaging, the gauge kinetic terms are
supplemented with the pieces needed to invert the kinetic operators
\cite{book}, and we get the well known gauge fixed action

\beqs
 S_0 + S_{fix} & = & {Tr \over 2} \int d x d^4 \th \( \psi \[ 
 - D^2 \bar{D}^2 - {1 \over \alpha} ( \bar{D}^2 D^2 - D \bar{D}^2 D ) 
 \] \bar{\psi} \right. \\ \nn \\
& & \left. + \; \bar{\psi} \[ - \bar{D}^2 D^2 - {1 \over \alpha} 
 (D^2 \bar{D}^2 - D \bar{D}^2 D )\] \psi + v \[ D \bar{D}^2 D - 
 {1 \over \alpha} ( D^2\bar{D}^2 + \bar{D}^2 D^2 ) \] v \)  .  \nn
\eeqs

\noindent
In $N=1$ superspace the physical prepotentials of the gauge multiplet
can be gauge fixed with apparently different $\alpha$ parameters. That 
is possible because in the usual formulation of the gauge multiplet 
the interactions of the potentials $\psi$ and $\bar{\psi}$ always involve 
the corresponding field strengths $\bar{\phi}$ and $\phi$. The 
propagator $\langle \psi (1) \bar{\psi} (2) \rangle$ is always acted upon 
with spinor derivatives from the interaction vertices to give 

\be 
 \langle D^2 \psi (1) \bar{D}^2 \bar{\psi} (2) \rangle = 
 \langle \bar{\phi}(1) \phi(2) \rangle 
\label{potential_prop}
\ee

\ni
or the conjugate expression. No matter what the value of $\alpha$ is, 
the derivatives in (\ref{potential_prop}) select the antichiral-chiral 
projector in the inverted kinetic operator

\be
 - {\bar{D}^2 D^2 \over \Box^2} - \a ( {D^2 \bar{D}^2 \over \Box^2 }
 - {D \bar{D}^2 D \over \Box^2} ) \ ,
\ee

\ni
and this is precisely the $\a$-independent piece. In the projective
superspace formulation of the gauge multiplet the interactions
(\ref{eq-Lagrangian2}) of the superfield $v_1$ do not involve the field
strength $\bar{\phi}$ and we have to be more careful. In $N=2$
superspace we use the gauge transformation (\ref{vector_transf}) to
fix the gauge for the whole real tropical multiplet containing $v_0$,
$v_{-1}$ and $v_1$. Their gauge transformations are not independent
and we must use the same gauge fixing parameter for both. Indeed, the
$N=2$ relations among coefficient superfields (\ref{proj_const})
guarantee that a single condition $v_2 = 0 = v_{-2}$ automatically
reproduces the $N=1$ gauge fixing conditions (\ref{n=1_gauge_fix}) on
the prepotentials $v_0$, $v_1$ and $v_{-1}$. However, it also gives
unwanted gauge fixing conditions on the prepotentials $v_3$, $v_4$,
$v_{-3}$ and $v_{-4}$ which are otherwise absent from the kinetic
action. We need additional {\em gauge fixings for the gauge fixing,
i.e.}  it is not enough to set $v_2 = 0 = v_{-2}$, but we have to fix
$v_n = 0 \; \forall \, n \neq -1,0,1$. This is precisely the $O(2)$
gauge discussed before.

 Our ultimate goal is to produce an invertible kinetic term for the
whole real vector multiplet with the full $N=2$ superspace measure, so
that we can compute the propagator with its full $N=2$ Grassmann
coordinate dependence. When we reduce this $N=2$ invertible kinetic
term to $N=1$ components we expect to find also a kinetic term for the
prepotentials $v_n , |n|>1$. These are unphysical degrees of freedom and
it maybe surprising to introduce kinetic operators for them. This is
however analogous to what happens with the unphysical fields of the
$N=1$ vector prepotential: although they can be set equal to zero in
Wess-Zumino gauge, the standard gauge fixing $D^2 v = 0 = \bar{D}^2 v$ 
gives an invertible kinetic term for the whole $v$.

\ni
With this argument in mind we tentatively propose the following 
$N=1$ gauge fixed action    

\beqs
\lefteqn{S_0 + S_{fix} = {1 \over 2} \int d x d^4 \th \; Tr \; \( 
 - {1 \over \a} \sum_{n \neq -1,0,1} v_n \Box v_{-n} + v_0 \left[ 
 D \bar{D}^2 D - {1 \over \alpha} ( D^2 \bar{D}^2 + \bar{D}^2 D^2 ) 
 \right] v_0  \  \right. \label{fixed_action} } \label{ans_fix} \\ \nn \\
& & \left. - v_1 \left[ D^2 \bar{D}^2 + {1 \over \a_{\ }} 
 ( \bar{D}^2 D^2 - D \bar{D}^2 D) \right]  v_{-1}
 - v_{-1} \left[ \bar{D}^2 D^2 + {1 \over \a} 
 ( D^2 \bar{D}^2 - D \bar{D}^2 D) \right] v_1 \) \nn \ . \qquad 
\eeqs

\ni
Of course this ansatz will only be justified once we have found a 
gauge fixing in $N=2$ superspace that reproduces this expression.

\section{$N=2$ propagator obtained from $N=1$ component propagators}

 We can now proceed to reconstruct the propagator of the vector
multiplet in $N=2$ superspace the same way it has been done for the
hypermultiplet \cite{massless_hyper}. We add to the action a
source term 

\be
S_j = \int d x d^4 \th \oint {d\zeta \over 2 \pi i \zeta} \; Tr \;
 j V = \int d x d^4 \th \sum_{n=-\infty}^{+\infty} Tr \; j_{-n} v_n \ ,
\label{eq-source}
\ee

\noindent 
where the source is a real tropical multiplet itself, making the whole
expression $N=2$ supersymmetric. The $\zeta$-coefficient sources are
unconstrained in $N=1$ superspace as we mentioned when we defined the
real tropical multiplet. This allows us to complete squares trivially 
on the $N=1$ component action of the vector multiplet. Integrating out
the $N=1$ superfields in the free theory path integral, we are left
with the following terms quadratic in sources

\beqs
 ln Z_0 [j, \bar{j}] & = & \; Tr \int d x d^4 \th \( - {1 \over 2} \:
  j_0 \left[ { D \bar{D}^2 D \over \Box^2} - \a {D^2\bar{D}^2 + 
 \bar{D}^2 D^2 \over \Box^2} \right] j_0 \right.   \\
& & \qquad \left. + \; j_1 \left[ {D^2\bar{D}^2  \over \Box^2} +  
 \a {\bar{D}^2 D^2 - D \bar{D}^2 D \over \Box^2} \right] j_{-1} + 
 \a \sum_{n=2}^{+\infty} j_n \Box^{-1} j_{-n} \) \nn \ .   
\eeqs

\ni
The vector multiplet propagator will have the following projection 
into $N=1$ superspace

\beqs
\langle V^a(1) V^b(2) \rangle_{|\theta_2 = 0} & = & \sum_{n=2}^{+\infty} 
 \langle v^a_{-n}(1) \; v^b_n(2) \rangle \( \zlZ \)^n + 
 \langle v^a_{-1}(1) \; v^b_1(2) \rangle \: \zlZ + 
 \langle v^a_0(1) \; v^b_0(2) \rangle  \nonumber       \\
& & + \langle v^a_1(1) \; v^b_{-1} (2) \rangle \: \zZl + 
 \sum_{n=2}^{+\infty} \langle v^a_n(1) \; v^b_{-n}(2) \rangle \( \zZl \)^n  \\
& = & \d^{ab} \( \a \sum_{n=2}^{+\infty} \left[ \( \zlZ \)^n + 
 \( \zZl \)^n \right] 
 \left[ D^2 \bar{D}^2 + \bar{D}^2 D^2 - D \bar{D}^2 D \right] \right. \nn \\
& & \; + \zlZ \left[ (1-\a) D^2\bar{D}^2  + \a (D^2\bar{D}^2 + 
                  \bar{D}^2 D^2 - D \bar{D}^2 D) \right]     \nn \\
& & \; + \left[- (1-\a) D \bar{D}^2 D + \a (D^2\bar{D}^2 + 
           \bar{D}^2 D^2 - D \bar{D}^2 D ) \right]    \nn \\ 
& & \left. \; + \zZl \left[ (1-\a) \bar{D}^2 D^2 + 
          \a (D^2 \bar{D}^2 + \bar{D}^2 D^2 - D \bar{D}^2 D) \right] \) 
 {\d^4 (x_{12}) \d^4 (\th_{12})\over \Box^2}  \nn \ . 
\eeqs

\ni
Rewriting the term proportional to $\a$ we obtain

\beqs
\langle V^a(1) V^b(2) \rangle_{|\th_2 = 0} & = & \d^{ab}  
 \( \a \sum_{n=-\infty}^{+\infty} \( \zlZ \)^n + (1-\a) \) \times 
 \label{gauge_prop-proj} \\
& & \qquad \qquad \qquad \times
 \left[ \zlZ D^2\bar{D}^2 - D \bar{D}^2 D + \zZl \bar{D}^2 D^2 \right]
 {\d^4 (x_{12}) \d^4 (\th_{12}) \over \Box^2}  \nn \ .
\eeqs

 When we quantized the hypermultiplet \cite{massless_hyper} we made 
the ansatz that the $N=2$ propagator of a projective multiplet should
contain the projective spinor derivatives $\na_1^4 \na_2^4 \d^8
(\th_1-\th_2)$ ($\na_1^4 = \na^2 (\zl) \bar{\na}^2 (\zl)$ ). The 
projection of this expression into $N=1$ superspace is

\be
 \na^4_1 \na^4_2 \d^8( \th_1 - \th_2 ) \mid_{\th_{\a}^2=0} 
 = (\zl  - \zZ)^2 ( \zl^2 \bar{D}^2 D^2 + \zZ^2 D^2 \bar{D}^2 
 - \zl \zZ D \bar{D}^2 D ) \d^4( \th_1 - \th_2 )\ . 
\label{nabla-delta}
\ee

\ni
Making the same ansatz for the tropical vector multiplet it is 
straightforward to realize that the $N=2$ propagator we look
for is

\be
 \langle V^a(1) V^b(2) \rangle =  
 \d^{ab} \( \a \sum_{n=-\infty}^{+\infty} \( \zlZ \)^n + (1 - \a) \)
 {\na_1^4 \na_2^4 \over \zl \zZ (\zZ - \zl)^2 \Box^2 } \d^8 (\th_{12})
 \d^4 (x_{12}) \ .
\label{2_vers_prop}
\ee

The first term is the only one present in Fermi-Feynman gauge $\a=1$. 
The infinite series defines the full (tropical) Dirac delta distribution 
\cite{massless_hyper} on the Riemann sphere for any function with a 
power series expansion in $\z$ 

\be
\oint {d\zl \over 2 \pi i \zl} F(\zl) 
 \d_{-\infty}^{+\infty} (\zZ, \zl) = \oint {d\zl \over 2 \pi i \zl} 
 F(\zl) \sum_{n=-\infty}^{+\infty} \( \zlZ \)^n = F(\zZ) \ .
\ee

\ni
Using the following identities 

\beqs
 {\na_1^4 \na_2^4 \over \zl \zZ (\zZ - \zl)^2 } & = & 
 {\bar{\na}_1^2 \na_1^2 \D_1^2 \bar{\na}_2^2 \over 4 \zl \zZ }  
 = {\na_1^4 \over \zl^2} \[ \zlZ D^2 \bar{D}^2 - D \bar{D}^2 D + 
 \zZl \bar{D}^2 D^2 \] \nn  \\  \nn \\
& = & {\na_1^2 \bar{\D}_2^2 \bar{\na}_2^2 \na_2^2 \over 4 \zl \zZ } 
 = \[ \zlZ D^2 \bar{D}^2 - D \bar{D}^2 D + \zZl \bar{D}^2 D^2 \] 
 {\na_2^4 \over \zZ^2}
\eeqs

\ni
and reordering 

\be
\sum_{n=-\infty}^{+\infty} \( \zlZ \)^n \[ {\zZ \over \zl} D^2 \bar{D}^2 - 
 D \bar{D}^2 D + {\zl \over \zZ} \bar{D}^2 D^2 \] = 
 \sum_{n=-\infty}^{+\infty} \( \zlZ \)^n \Box \ ,
\ee

\ni
we can rewrite the $N=2$ propagator in Fermi-Feynman gauge $\a=1$ in two 
equivalent forms 

\beqs
 \langle V^a(1) V^b(2) \rangle & = & \d^{ab} {\nabla_1^4 \over \zeta_1^2 \Box} 
 \delta^8 (\theta_1 - \theta_2) \delta^4 (x_1 - x_2) 
 \sum_{n=-\infty}^{+\infty} \( \zlZ \)^n  \label{form1} \\
 & = & \d^{ab} {\nabla_2^4 \over \zeta_2^2 \Box} 
 \delta^8 (\theta_2 - \theta_1) \delta^4 (x_1 - x_2) 
 \sum_{n=-\infty}^{+\infty} \( \zlZ \)^n  \ . \label{form2}
\eeqs

In Landau gauge $\a=0$ the gauge fixing is given an infinite weight in
the path integral. Not surprisingly, the propagator is the 
one corresponding to an $O(2)$ multiplet \cite{massless_hyper} 
$V^{O(2)} = \eta / \z$ with kinetic term

\be
 S_0 = - \int d x d^4 \th \oint { d \z \over 2 \pi i \z} {\eta \over \z}
 \Box {\eta \over \z} \ .
\ee

\section{Gauge fixing in $N=2$ superspace}

 The $N=2$ vector multiplet has a well defined kinetic action in
$N=2$ superspace as we have seen, and the $\zeta$-coefficient superfields 
can be suitably gauge fixed to produce invertible kinetic terms in 
$N=1$ superspace. It is therefore natural to expect that a $N=2$ gauge
fixing term exists, allowing us to invert the $N=2$ kinetic operator. 

 We want to gauge fix the vector multiplet in the $N=2$ free theory
path integral 

\be
 Z_0 =  \int D [V] D [f] \; \Delta^{-1}_{FP} \; \d (V_G-f) e^{ i S_0(V)} 
 e^{i S_{avg}(f)} \ ,
\ee

\ni
where we define the truncated tropical fields $V_G$ and $f$ using the
polar Dirac delta distributions on the Riemann sphere 
\cite{massless_hyper}
 
\be
 V_G (\z) = \oint {d \z' \over 2 \pi i \z'} \( \d_{(- \infty)}^{(-2)} 
 (\z, \z') + \d_{(2)}^{(+ \infty)} (\z, \z') \) V (\z') = 
 \sum_{n=-\infty}^{-2} \z^n v_n  + 
 \sum_{n=2}^{+\infty} \z^n v_n \ .
\ee

 In the abelian case the Faddeev-Popov ghosts only have a quadratic 
kinetic term and since they decouple from the other fields we will 
not be concerned with them anymore. In the nonabelian case we obtain
the Fadeev-Popov determinant as usual from the functional derivative
of the gauge fixing function

\beqs
 { \d V_G (1) \over \d (\Lambda (2), \bar{\Lambda}(2) ) } & = &
 \oint {d \z_0 \over 2 \pi i \z_0} \( \d_{(- \infty)}^{(-2)} (\zl,\z_0) 
 + \d_{(2)}^{(+ \infty)} (\zl,\z_0) \) \times \\
& & \times L_{V (\z_0,\th_1,x_1) \over 2} \[ \( \d_{(0)}^{(+ \infty)} 
 (\z_0,\zZ) \na_0^4 + \d_{(- \infty)}^{(0)} (\z_0,\zZ) 
 {\na_0^4 \over \z_0^4} \) + \right. \nn \\
& & \left. + \; coth L_{V ( \z_0, \th_1, x_1) \over 2} 
 \( \d_{(0)}^{(+ \infty)} (\z_0,\zZ) \na_0^4 - 
 \d_{(- \infty)}^{(0)} (\z_0,\zZ) {\na_0^4 \over \z_0^4} \) \] 
 \d^8 (\th_{12}) \d^4 (x_{12}) \ ,  \nn
\eeqs

\ni
where we have used the functional derivatives with respect to
(ant)arctic multiplets $ \Lambda = \na^4 \Psi$ and 
$\bar{\Lambda} = \na^4 \bar{\Psi} / \z^4$ \cite{massless_hyper}.
The inverse Faddeev-Popov determinant is obtained by taking the
matrix elements of this operator evaluated on the infinite 
basis of unconstrained anticommuting real tropical superfield points 
$b(\zl, \th_1, x_1)$, $ c(\zZ, \th_2, x_2)$, and integrating out 
its exponential in the path integral

\beqs
 \D^{-1}_{FP} = \int D[b] D [c] exp \( Tr \int d x_1 d x_2 d^8 \th_1
 d^8 \th_2 \oint {d\z_0 \over 2 \pi i \z_0} {d\zl \over 2 \pi i \zl} 
 {d\zZ \over 2 \pi i \zZ} b (\zl, \th_1, x_1) \right. & & \nn \\ 
 \left. \( \d_{(- \infty)}^{(-2)} (\zl, \z_0) + 
 \d_{(2)}^{(+ \infty)} (\zl, \z_0) \)
 { \d V (\z_0, \th_1, x_1) \over \d (\Lambda (2), \bar{\Lambda}(2) ) }
 c(2) \) & . &
\eeqs

\ni
Defining the (ant)arctic ghost multiplets

\beqs
 C = \na_0^4 \sum_{n=0}^{+\infty} \z_0^n c_n & , & 
 \bar{C} = {\na_0^4 \over \z_0^4} \sum_{n=0}^{+\infty} {\bar{c}_n \over
 (-\z_0)^n} \ , \\
 B = {\na_0^4 \over \z_0^2} \sum_{n=2}^{+\infty} \z_0^n b_n & , &
 \bar{B} = {\na_0^4 \over \z_0^2} \sum_{n=2}^{+\infty} 
 {\bar{b}_n \over (-\z_0)^n} \ , \nn
\eeqs

\ni
the resulting $N=2$ ghost action is
 
\be
 S_{FP} = \int d x d^4 \th \oint {d \z_0 \over 2 \pi i \z_0}
 (B + \bar{B}) L_{V \over 2} [ (C + \bar{C}) + coth L_{V \over 2} 
 (C- \bar{C})] \ .
\ee

\ni
The kinetic piece in this action is 

\be
S_0^{FP} = \int d x d^4 \theta \oint {d \z_0 \over 2 \pi i \z_0} \;
 Tr \;(B + \bar{B}) (C - \bar{C}) = 
  \int d x d^4 \theta \oint {d \z_0 \over 2 \pi i \z_0} \; Tr \;
 (\bar{B} C - B \bar{C}) \ ,
\ee

\ni 
and the ghost quantization is very similar to that of two mixed (ant)arctic
hypermultiplets in the adjoint representation of the gauge group, with
the peculiarity that this superfields are anticommuting. Performing
the contour integral we obtain $N=1$ chiral and and complex linear
kinetic ghost terms of the form (\ref{n1_ghosts}).

 The gauge fixing weight we propose is the following  

\be
 S_{avg} = - {Tr \over 2 \a } \int d x d^8 \th \( 
 \oint_{|\zl| < |\zZ|} {d \zl \over 2 \pi i} {d\zZ \over 2 \pi i} 
 \: { \zl f(\zl) f(\zZ) \over (\zZ- \zl)^3} 
 + \oint_{|\zZ| < |\zl|} {d \zl \over 2 \pi i} {d\zZ \over 2 \pi i} 
 \: { \zZ f(\zl) f(\zZ) \over (\zl- \zZ)^3} \) \ . 
\ee

\ni
Integrating the weighted function $f$ in the path integral we obtain 
the gauge fixing action $S_{fix} = S_{avg} (f \rightarrow V_G)$. 
Performing the contour integrals we notice that we may replace 
$V_G \rightarrow V$ because the integration automatically projects 
out the components $v_{-1}, v_0$ and $v_1$

\beqs
S_{fix} & = & - {Tr \over 2 \a } \int d x d^8 \th \( \oint_{|\zl| < |\zZ|} 
 {d \zl \over 2 \pi i} {d\zZ \over 2 \pi i} \:  
 { \zl V (\zl) V (\zZ) \over (\zZ- \zl)^3} + 
 \oint_{|\zZ| < |\zl|} {d \zl \over 2 \pi i} {d\zZ \over 2 \pi i} \: 
 { \zZ V (\zl) V (\zZ) \over (\zl- \zZ)^3} \)  \nn    \\ \nn \\
& = & - {1 \over \alpha} \int d x d^8 \th  
 \sum_{n=2}^{+\infty} {n (n-1) \over 2} \; Tr \; v_{-n} v_n  \ .
\label{gauge_fixing}   
\eeqs

\ni
When we project this expression into $N=1$ superspace we find the 
extra pieces needed to invert the kinetic terms. A convenient way to
perform such projection is to replace the $N=2$ superspace measure by
$D^2 \bar{D}^2 Q^2 \bar{Q}^2 = D^2 \bar{D}^2 \na^4 / \z^2$ and act
with the projective derivatives on the integrand \cite{massless_hyper}. 
To simplify the contour integrals we use the identity

\be
 { \na_1^4 \over (\zZ-\zl)^2 } V (\zZ) = \na_1^2 \bar{D}^2 V(\zZ) \ ,
\label{double_pole_id}
\ee

\ni
and we rewrite the $N=1$ projection of the gauge fixing action as 
follows

\beqs
 S_{fix} = - {Tr \over 2 \a} \int d x D^2 \bar{D}^2 \( 
 \oint_{|\zl| < |\zZ|} {d\zl \over 2 \pi i} {d\zZ \over 2 \pi i} 
 V (\zl) {\na_1^2 \over \zl^2} \bar{D}^2   
 {\zl \over (\zZ-\zl)} V (\zZ) \right.  \nn \\   \nn \\
 \left. + \oint_{|\zZ| < |\zl|} {d\zl \over 2 \pi i} {d\zZ \over 2 \pi i} 
 V (\zl) {\na_1^2 \over \zl^2} \bar{D}^2  
 {\zZ \over (\zl-\zZ)}  V (\zZ) \) & . &
\eeqs

\ni
We have transformed a triple pole into a simple one. The remaining
pole can be written as a convergent geometric series  

\beqs
 S_{fix} =  - {Tr \over 2 \a} \int d x d^4 \th \( 
 \oint_{|\zl| < |\zZ|} {d\zl \over 2 \pi i} {d\zZ \over 2 \pi i} 
 V (\zl) {\na_1^2 \over \zl^2} \bar{D}^2  
 \zZl \sum_{n=0}^{+\infty} \( \zZl \)^n V (\zZ) \right. & & \nn \\ \nn \\
 \left.  + \oint_{|\zZ| < |\zl|} {d\zl \over 2 \pi i} 
 {d\zZ \over 2 \pi i} V (\zl) {\na_1^2 \over \zl^2} 
 \bar{D}^2 \zlZ \sum_{n=0}^{+\infty} \( \zlZ \)^n   
 V(\zZ) \) & . &
\eeqs

\ni
In this form the only poles are in the origin of the complex plane 
and the radial ordering is irrelevant for the contour integration. 
We can deform the contours so that they become overlapping and combine
both terms in the same double integral. We also use the identity
(\ref{double_pole_id}) in the $N=1$ projection of the kinetic action 
(\ref{n2_vector_action}), and we find the following $N=1$ gauge fixed 
action 

\be
 S_0 + S_{fix} =  - {Tr \over 2} \int d x d^4 \th \oint 
 {d\zl \over 2 \pi i \zl} {d\zZ \over 2 \pi i \zZ} V (\zl) 
 {\na_1^2 \over \zl^2} \bar{D}^2 \zl \zZ \left[ 
 1 + {1 \over \a} \sum_{n \neq 0} \( \zlZ \)^n \right] V (\zZ)  \ .
\label{origin_pole} 
\ee

\ni
Finally, we rewrite the differential operator acting on $V(\zZ)$ 

\be
 \( \zlZ \)^{n+1} \na_1^2 \bar{D}^2 V(\zZ) = 
 \[ \( \zlZ \)^{n+1} D^2 \bar{D}^2 - \( \zlZ \)^n D \bar{D}^2 D + 
 \( \zlZ \)^{n-1} \bar{D}^2 D^2 \] V(\zZ) \ ,
\ee 

\ni
and we recover the form of the $N=1$ gauge fixed action
(\ref{ans_fix}) we proposed 

\beqs
 S_0 + S_{fix} & = & - {Tr \over 2} \int d x d^4 \th \oint 
 {d\zl \over 2 \pi i \zl} {d\zZ \over 2 \pi i \zZ} V (\zl) 
 \( \zZl \[ \bar{D}^2 D^2 + {1 \over \a} ( D^2 \bar{D}^2 - 
  D \bar{D}^2 D) \] \right. \nn \\ \nn \\
& & \qquad + \[ {1 \over \alpha} ( D^2 \bar{D}^2 + \bar{D}^2 D^2 ) 
 - D \bar{D}^2 D \] + \zlZ \[ D^2 \bar{D}^2 + {1 \over \a} 
 ( \bar{D}^2 D^2 - D \bar{D}^2 D) \] \nn \\
& & \left. \qquad + {1 \over \a} \sum_{n \neq -1,0,1} \( \zZl \)^n  
 \[ \bar{D}^2 D^2 + D^2 \bar{D}^2 - D \bar{D}^2 D \] \) 
 V (\zZ) \ . 
\label{n1_fixed_action}
\eeqs

\section{Computation of the $N=2$ propagator in $N=2$ superspace}

We have successfully found a gauge-fixing in $N=2$ superspace that 
reproduces the ansatz (\ref{fixed_action}) for a gauge fixed action 
in $N=1$ components. To invert the kinetic term in $N=2$ superspace 
we use an unconstrained real tropical superfield $X (\z)$ which
defines a {\em prepotential} for the projective gauge multiplet 

\be
 V (\z) = { \nabla^4 \over \z^2} X (\z) \ .
\ee

\ni
We rewrite the gauge fixed action in $N=2$ superspace as 

\beqs
 \lefteqn{ S_0 + S_{fix} = - {Tr \over 2} \int d x d^8 \th \oint 
 {d\zl \over 2 \pi i \zl} {d\zZ \over 2 \pi i \zZ} X (\zl) 
 {\bf K} (\zl, \zZ) X (\zZ) } \\
& = & - {Tr \over 2} \int d x d^8 \th \oint 
 {d\zl \over 2 \pi i \zl} {d\zZ \over 2 \pi i \zZ} X (\zl) 
 {\na_1^2 \over \zl^2} \bar{D}^2 \zl \zZ  \left[ 
 1 + {1 \over \a} \sum_{n \neq 0} \( \zlZ \)^n \right] 
 {\na^4 (\zZ) \over \zZ^2} X (\zZ) \nn  \ .
\label{n2_fixed_act}
\eeqs

Now we must add a source term in $N=2$ superspace
that reproduces the $N=1$ sources (\ref{eq-source}). The source term 
we want must involve and unconstrained real tropical source $J$, so 
that the projection into $N=1$ superspace contains the projective 
multiplets $V$ and $j = \na^4 J / \z^2$

\be
 S_J = \int d^4x d^8 \th \oint {d\z \over 2 \pi i \z} \; Tr \; 
 J (\z) V(\z) = \int d^4x D^2 \bar{D}^2 \oint {d\z \over 2 \pi i \z} 
 \; Tr \; j(\z) V (\z) \ .
\ee 

 To integrate out the gauge field $X(\z)$ in the $N=2$ path integral we 
have to complete the squares in the fixed free action with sources. 
This requires inserting a projector operator in the source term in 
such a way that we can factor the $N=2$ kinetic operator. Then we can 
redefine the unconstrained tropical superfield $X(\z)$ by a shift

\beqs
 S_0 + S_{fix} + S_J & = & - {Tr \over 2} \int d x d^8 \th \oint 
 {d\zl \over 2 \pi i \zl} {d\zZ \over 2 \pi i \zZ} \[ X (\zl) + 
 {\cal J}(\zl) \] {\bf K} (\zl, \zZ) \[ X (\zZ) + {\cal J}(\zZ) \] \nn \\  
& & + {Tr \over 2} \int d x d^8 \th \oint {d\zZ \over 2 \pi i \zZ} 
 J(\zZ) {\na^4 (\zZ) \over \zZ^2} {\cal J}(\zZ) \ , 
\label{pi_exponent}
\eeqs

\ni
where the shift superfield

\be
 {\cal J} (\zl) = \oint {d \z_0 \over 2 \pi i \z_0} {\bf P} (\zl, \z_0) 
 J (\z_0)
\ee

\ni
obeys

\beqs
 S_J & = & \int dx d^8 \oint {d\zl \over 2 \pi i \zl} J (\zZ) 
 {\na^4 (\zZ) \over \zZ^2} X(\zZ)  \\
& = & \int dx d^8 \th \oint {d\zl \over 2 \pi i \zl} 
 {d\zZ \over 2 \pi i \zZ} \[\oint {d \z_0 \over 2 \pi i \z_0}  
 {\bf P} (\zl, \z_0) J (\z_0) \] {\bf K} (\zl, \zZ) X (\zZ) \nn \ .
\label{proj_cond}
\eeqs

\ni 
The last term in (\ref{pi_exponent}) gives the source dependence of
the free theory path integral. We expect it to contain the
differential operators present in the propagator (\ref{2_vers_prop}),
and therefore ${\bf P} \propto \na^4$. Indeed acting with 
additional projective derivatives on the kinetic operator we obtain

\be
 \na_0^4 \( \na_1^2 \bar{D}^2 \bar{\na}_2^2 \na_2^2 \) 
 = \na_0^2 \bar{\na}_0^2 \( \na_1^2 \bar{D}^2 \bar{Q}^2 \na_2^2 \)
 = (\z_0 - \zl)^2 \Box \na_0^2 \bar{D}^2 \bar{Q}^2 \na_2^2 \ ,
\ee
 
\ni
and the operator we are looking for is  

\be
 {\bf P} (\z_0, \zl) = \( 1 + \a \sum_{n \neq 0} 
 \( {\zl \over \z_0} \)^n \) {\zl \na_0^4 \over \z_0 (\z_0-\zl)^2 \Box^2 } 
 = \overline{\bf P} (\z_0, \zl) \ .
\ee
 
\ni
It is very straightforward to check that it obeys the condition we imposed
in (\ref{proj_cond})

\be
 \oint {d \zl \over 2 \pi i \zl} {d \z_0 \over 2 \pi i \z_0}
 J (\z_0) {\bf P} (\z_0, \zl) {\bf K} (\zl, \zZ) =  
 \oint {d\z_0 \over 2 \pi i \z_0} \sum_{n= -\infty}^{+\infty} 
 \( {\zZ \over \z_0} \)^n J(\z_0) 
 { \na_0^2 \bar{D}^2 \bar{Q}^2 \na_2^2 \over \zZ^2 \Box} = 
 J(\zZ) {\na_2^4 \over \zZ^2} \ .
\ee 

\ni
Thus the free theory path integral is 

\beqs
 ln Z_0 [J] & = & {Tr \over 2} \int d x d^8 \th \oint 
 {d\z' \over 2 \pi i \z'} {d\z \over 2 \pi i \z} J(\z') 
 {\na^4 (\z') \over \z'^2} {\bf P} (\z', \z) J (\z) \ , \\
& = & {Tr \over 2} \int d x d^8 \th \oint {d\z' \over 2 \pi i \z'} 
 {d\z \over 2 \pi i \z} J(\z') 
 \( \a \sum_{n= -\infty}^{+\infty} \( {\z \over \z'} \)^n + (1 -\a) \)
 {\na^4 (\z') \na^4 (\z) \over \z' \z (\z - \z')^2 \Box^2 } J (\z) \nn
 \ .
\eeqs

\ni 
Now the propagator in $N=2$ superspace can be simply
obtained by functionally differentiating the path integral with
respect to the sources. Since the source is unconstrained, the 
functional derivative with respect to it is just the product of Dirac 
delta distributions in $N=2$ superspace and in the Riemann sphere. The
propagator we find reproduces correctly our ansatz (\ref{2_vers_prop})

\beqs
 \langle V^a (1) V^b (2) \rangle & = & {\d \over \d J_a(1)} 
 {\d \over \d J_b (2) } Z_0 [J]   \\
& = & \d^{ab} \( \a \sum_{n=-\infty}^{+\infty} \( \zlZ \)^n + (1 - \a) \)
 {\na_1^4 \na_2^4 \over \zl \zZ (\zZ - \zl)^2 \Box^2 } \d^8 (\th_{12})
 \d^4 (x_{12}) \ . \nn
\eeqs

\section{Vertices}

 The abelian vector multiplet can only interact with charged
fields. The best example we have of such interactions is the charged
(ant)arctic hypermultiplet coupling (\ref{eq-Lagrangian2}). This
couplings can be generalized to nonabelian vector multiplets
interacting with hypermultiplets in the fundamental or adjoint
representation of the gauge group. In addition the $N=2$ superpotential 
of the nonabelian vector multiplet contains self-interactions. At the
time of writing this manuscript we have not yet been able to rewrite
them as functionals of the real tropical multiplet, as we did
with the kinetic term. We can only give at this moment the vertex
factors of the first type. 

As we explained in \cite{massless_hyper}, when the propagators of 
projective multiplets contain the maximum number of projective 
derivatives we can formally put such derivatives in the vertex
factors. The factor corresponding to an interaction 
(\ref{eq-Lagrangian2}) where all fields give internal lines 
will be

\be
  \int d^8 \theta  \oint {d \z \over 2 \pi i \z} \na^4  
 exp \( \na^4 \) \na^4 \ .
\label{vertex_factor}
\ee 

\ni
whereas in interactions with external and internal fields we only 
replace the latter with projective derivatives.

\section{Feynman rules for the interacting vector multiplet and
hypermultiplet}

The Feynman rules for construction of diagrams in $N=2$ superspace are
a simple generalization of those given in \cite{massless_hyper}. As we
mentioned above we can choose to put the projective derivatives of the
propagators in the corresponding lines of the interaction vertices, 
working formally with propagators  

\be
 \langle \U (1) \bU (2) \rangle = (-) \d_{(0)}^{(+\infty)} (\zl, \zZ) 
 { 1 \over \zZ^2 ( \zl - \zZ )^2 \Box } \d^8( \th_1 - \th_2 ) 
 \d (x_1-x_2) \ ,
\ee

\ni
and

\be
 \langle V^a(1) V^b(2) \rangle = \d^{ab} \( \a \sum_{n=-\infty}^{+\infty} 
 \( \zlZ \)^n + (1 - \a) \) {1 \over \zl \zZ (\zZ - \zl)^2 \Box^2 } 
 \d^8 (\th_{12}) \d^4 (x_{12}) \ .
\ee

However, if we use the simplified form of the gauge propagator in
Fermi-Feynman gauge (\ref{form1}) such a choice is not possible because
it does not have enough projective derivatives. In many cases it is much 
simpler to use this propagator, and to treat all multiplets the 
same way we keep the spinor derivatives in the propagators. 
In that case we work with hypermultiplet propagators 

\be
 \langle \U (1) \bU (2) \rangle = (-) \d_{(0)}^{(+\infty)} (\zl, \zZ) 
 { \na_1^4 \na_2^4 \over \zZ^2 ( \zl - \zZ )^2 \Box } \d^8( \th_1 - \th_2 ) 
 \d (x_1-x_2) \ .
\ee

\ni
and vector multiplet propagators

\beqs
 \langle V^a(1) V^b(2) \rangle & = & \d^{ab} {\nabla_1^4 \over \zeta_1^2 \Box} 
 \delta^8 (\theta_1 - \theta_2) \delta^4 (x_1 - x_2) 
 \sum_{n=-\infty}^{+\infty} \( \zlZ \)^n  \nn \\
 & = & \d^{ab} {\nabla_2^4 \over \zeta_2^2 \Box} 
 \delta^8 (\theta_2 - \theta_1) \delta^4 (x_1 - x_2) 
 \sum_{n=-\infty}^{+\infty} \( \zlZ \)^n  
\eeqs

 To construct a given diagram we expand the exponential of the
interacting action and we let the functional derivatives with respect 
to internal line sources act on the free theory path integral. This 
standard procedure gives the combinatorial factors associated with 
each diagram.

After this we follow the usual strategy in the computation of
superspace diagrams: we extract a total derivative $\na_i^4$ for each
vertex $i$ and complete the restricted measure $d^4 \th_i$ to a full
$N=2$ superspace measure $ \z_i^2 d^8 \th_i$. Then we reduce the
Grassmann coordinate dependence of all propagators except the last one
to bare Dirac delta functions.

In Fermi-Feynman gauge
the most efficient way to perform these two steps is to extract the
total derivative $\na_i^4$, used to complete the measure of the vertex
$i$, from a gauge internal line connecting it to a vertex $j$. This
manipulation reduces the gauge propagator to a Grassmann delta
function $\d^8 (\th_{ij})$ and a Riemann sphere delta distribution
$\d_{(-\infty)}^{(+\infty)} (\z_i, \z_j)$. Integrating with the $N=2$
superspace measure and the complex contour measure of the vertex $i$,
we bring the vertices $i$ and $j$ to the same point in $N=2$ Grassmann
space and in the Riemann sphere. After reducing all the gauge
propagators in this fashion, we complete the superspace
measure in all other vertices by extracting total derivatives from the
hypermultiplet propagators. The final steps are the same as those
described in \cite{massless_hyper}: we perform the ``D''-algebra using
transfer rules and integration by parts to reduce the hypermultiplet
propagators to bare Grassmann delta functions. Integrating the
corresponding coordinates, at the end of the process we are left with
one bare Grassmann delta function multiplying certain number of spinor
derivatives that act on a similar delta function

\be
 \d^8 (\th_n - \th_m) \na_p \dots \na_q \d^8 (\th_n - \th_m) \ .
\ee
  
Any number of spinor derivatives larger than $8$ must be reduced using
the anticommutation relations (\ref{algebra_id}) of projective 
derivatives. Any number of spinor derivatives smaller than $8$ makes 
the product
vanish, while $8$ derivatives completely eliminate the Dirac delta
function on the right \cite{massless_hyper}. The last bare delta 
function can be integrated over one of the Grassmann coordinates and 
the final amplitude is local Grassmann functional integrated with the 
full $N=2$ superspace measure. 

 Finally, we must perform the complex contour integrals. As we
mentioned in references \cite{massless_hyper} and \cite{eff_action}
this step usually involves using a radial ordering prescription on the
complex contours. Some examples of 1-loop and 2-loop calculations
using these Feynman rules will be presented in a future publication
\cite{eff_hyper}.

\section{Acknowledgments}
The author would like to thank Martin Ro\v{c}ek and R. von Unge for helpful
discussions. This work was partially supported by NSF grant No Phy 9722101.

\end{document}